\documentclass[a4paper,11pt]{article}
\usepackage{aaskaiid}
\usepackage{orcidlink}
\usepackage{aas_macros}

\title{ Probing High-redshift Intracluster Medium Using SKA}
\ShortTitle{El-Gordo with SKA}

\author[1, 2]{Ramananda Santra \orcidlink{0009-0002-0373-570X}}
\ShortName{Santra et al.} 
\author[1]{Ruta Kale\orcidlink{0000-0003-1449-3718}}

\affiliation[1]{National Centre for Radio Astrophysics, Tata Institute of Fundamental Research, S. P. Pune University Campus, Ganeshkhind, Pune 411007, India}

\affiliation[2]{International Centre for Theoretical Sciences, Tata Institute of Fundamental Research, Bangalore 560089, India}

\emailAdd{ramananda1999@gmail.com}
\emailAdd{ruta@ncra.tifr.res.in}

\abstract{Galaxy clusters host vast reservoirs of magnetised plasma in their intracluster medium (ICM), where turbulence and shocks generated during structure formation sometimes give rise to diffuse synchrotron emission in the form of radio halos and relics. However, the origin and amplification of the magnetic field in the ICM remains among the least understood problems, particularly at high redshift, where direct detections are scarce. Recent observations of clusters at z $\gtrsim 0.6$ have revealed radio powers of the diffuse sources comparable to those of nearby systems, implying that efficient magnetic amplification suppressed inverse Compton losses. Such rapid growth of the magnetic fields after a few Gyrs of the Big Bang challenges standard dynamo timescales, and it raises fundamental questions about the mechanisms that strengthened cluster magnetic fields. In this chapter, we highlight the massive merging system El Gordo (ACT-CL J0102$-$4915, z$\sim$0.87) as a unique testbed for investigating theoretical models under extreme conditions. Proposing as a SKA science verification target, El Gordo will allow detailed studies of the ICM magnetic field structure, strength through broadband continuum and polarisation measurements with SKA-Low and SKA-Mid, across staged deployments starting from AA0.5 to AA4, reaching a physical resolution of $\sim$ 20-30 kpc at our redshift of interest. We further outline how upcoming SKA surveys will build statistical samples of high-redshift clusters, enabling tests of different models of magnetic fields and cosmic rays in the formation of large-scale structure.
}

\begin{document}
\maketitle

\section{Introduction}

Magnetic fields are a fundamental but elusive component of the universe, influencing structures from galaxies to the large-scale cosmic web \citep[see][for reviews]{ryu11, subhra}. In galaxy clusters, the intracluster medium (ICM) hosts magnetic fields that regulate plasma microphysics, control heat transport, and shape non-thermal radio emission \citep[see][for reviews]{bru14, 2024NatCo..15.1006H}. The non-thermal radio emission in galaxy clusters is broadly classified into two main categories \citep{fer12}: radio halos, centrally located steep-spectrum ($\alpha \lesssim -1.0$; we use S $\propto \nu^{\alpha}$, where $\alpha$ is the spectral index) unpolarised sources; and radio relics, which trace outgoing merger shocks and exhibit elongated morphology with high linear polarisation fraction \citep[see][for reviews]{wee19, 2023JApA...44...38P}. At low redshift ($z < 0.5$), observations with LOFAR, GMRT, MeerKAT, and VLA have revealed diffuse synchrotron sources extending up to 2~Mpc or beyond, tracing turbulence and magnetic fields in the ICM on scales comparable to the cluster virial radius \citep[e.g.,][]{kal13, kal15, knowles22, 2022A&A...660A..78B}. These imply magnetic field strengths of a few~$\mu$G, pervasive across much of the cluster volume and consistent with turbulent amplification. However, the origin and amplification of these fields remain poorly understood, especially at high redshift \citep[e.g.,][]{sub06, lehle25}. Beyond $z \gtrsim 0.6$, only a few clusters show confirmed diffuse emission, owing to steep spectra, surface-brightness dimming, and limited sensitivity. This gap limits our understanding of when and how cluster-scale magnetic fields emerged and evolved. Recent SKA-precursor results have begun to bridge this divide \citep[e.g.,][]{2014ApJ...786...49L, digen21, phuravhathu25}. Observations of $\mu$G-level fields at $z \sim 0.8$ suggest amplification within the first $\sim$3–4~Gyr after the Big Bang, also Faraday rotation studies reveal highly ordered fields at similar epochs. These findings challenge models predicting slower dynamo growth and call for reassessment of early cluster magnetisation \citep[e.g.,][]{beresnyak16, tevlin25}.

Magnetic fields in clusters likely arise from weak seed fields ($B \sim 10^{-18}$–$10^{-15}$~G) that are amplified by merger- and accretion-driven turbulence \citep[e.g.,][]{seta20, 2023ApJ...944..100M, kriel25}. The seeds may be primordial, generated during inflation, phase transitions, or astrophysical, injected by supernovae, galactic winds, and AGN outflows \citep[see][for reviews]{widrow2012, durrer13}. In the absence of large-scale rotation, amplification proceeds via a small-scale turbulent dynamo, where chaotic ICM motions stretch and fold field lines until saturation \citep[e.g.,][]{brandenburg04, 2013MNRAS.432..668L}. Cosmological simulations support this picture, demonstrating amplification to $\mu$G strengths within a few Gyr and producing filamentary structures associated with shocks and cold fronts \citep[e.g.,][]{miniati15, vaz16, donnert18}. Yet direct observational tests at high redshift remain scarce due to low surface brightness, contamination from radio galaxies, and limited angular resolution \citep[e.g.,][]{digen21, digennaro25}. Therefore, understanding magnetic-field evolution in high-$z$ clusters is crucial because they:

\begin{itemize}
\item Regulate the ICM’s thermal balance and energy transport;
\item Govern non-thermal phenomena such as halos, relics, and AGN feedback; and
\item Reveal how and when the universe became magnetised on large scales.
\end{itemize}

Key open questions include: How are seed fields generated and amplified? What processes dominate their growth? How do field strength, coherence, and morphology evolve with cluster dynamics and redshift?

The Square Kilometre Array Observatory (SKAO) is going to be the largest radio telescope that has been built to date. The unprecedented sensitivity and sharp angular resolution will revolutionise radio astronomy \citep[e.g.,][]{2009IEEEP..97.1482D}. In this chapter, we will focus on the expected impact of the SKA on high-redshift cluster science, highlighting El Gordo (ACT-CL~J0102$-$4915, $z=0.87$) as a testbed system (science verification target) to probe magnetic field amplification and plasma physics behind particle acceleration during the epoch of cluster assembly. The chapter is organised as follows: Sec~\ref{obs} introduces the current understanding of the observational status at high redshift, and Sec~\ref{theroy} reviews the possible mechanisms of amplification of seed magnetic fields, which should be tested observationally. An overview of the target cluster is provided in Sec~\ref{el-gordo}. The expected SKAO sensitivities for the various arrays and the prospects are provided in Sec~\ref{ska}. A simple estimation of the detection threshold is presented in Sec~\ref{population}. Finally, the summary is presented in Sec~\ref{summ}.

\section{Observational lessons of high-redshift clusters}\label{obs}

Until recently, confirmations of diffuse synchrotron emission (halos and relics) in clusters at z $\gtrsim$ 0.6 were scarce, leaving a gap in our understanding of the non-thermal ICM at those epochs. At low redshifts (z $<$ 0.5), facilities such as the JVLA and GMRT have identified numerous radio halos and relics, indicating magnetic field strengths of a few $\mu$G in these systems \citep[e.g.,][]{ven08,bon14,kal15,cuc21, 2021PASA...38...10D,2022MNRAS.514.5969K, 2024PASA...41...26D}, supporting turbulent-amplification models for ICM magnetic fields. In contrast, high-redshift observations are hampered by steep synchrotron spectra, cosmological surface-brightness dimming, and severe inverse Compton losses, as (1 + z)$^{4}$ increases rapidly with redshift. Recent sensitivity upgrades of SKA precursors, including MeerKAT, are now overcoming these limitations \citep[e.g.,][]{knowles21, phuravhathu25}. In a MeerKAT survey of six massive clusters at 1.01 $<$ z $<$ 1.31, diffuse emission was confidently detected in four clusters, with tentative signals in the other two \citep[e.g.,][]{knowles21}. The radio powers span $\sim$ (0.46 $-$ 4.5) $\times$ 10$^{24}$ W~Hz$^{-1}$, with linear extents of 0.47$-$1.08 Mpc. Additionally, a radio halo was discovered in ACT-CL J0329.2$-$2330 at z = 1.23 (M$_{500}$ = 9.7 $\times$ 10$^{14}$ M$_{\odot}$) with a spectral index of $-$1.5 \citep{sikhosana25}. These results show that $\mu$G-level magnetic fields and non-thermal emission were already in place at early cosmic epochs \citep[e.g.,][]{2019ApJ...881L..18C,digen21}.

This picture has further evolved with the advent of upgraded low-frequency facilities such as LOFAR. The improved sensitivity and dynamic range have pushed the detection threshold for diffuse emission to new limits. The LOFAR Two-Metre Sky Survey (LoTSS-DR2; \citealt{shimwell22}), with a typical noise level of $\sim$ 80$\mu$Jy~beam$^{-1}$ at 6$''$ resolution, has enabled systematic searches for cluster-scale synchrotron emission out to $z \sim 1.2$. A key milestone comes from the Massive and Distant Clusters of WISE Survey (MaDCoWS) analysis in LoTSS-DR2 \citep[e.g.,][]{digennaro25}. This study investigated 56 clusters in the redshift range $0.78 < z < 1.53$, detecting extended ($350$–$500$ kpc) diffuse emission in five systems—some at $z > 1.2$. These candidate halos occupy the same or higher locus of the radio power–mass (P$_\nu$–M$_{500}$) relation known from lower-redshift clusters, implying an efficient particle acceleration process when the universe was less than half its current age. Despite enhanced radiation losses from inverse Compton scattering, these results suggest that non-thermal acceleration processes remain highly efficient \citep{digen21}. Similarly, detections of diffuse emission in ACT-CL J0329.2$-$2330 ($z = 1.23$; \citealt{sikhosana25}) and other Planck- and ACT-selected systems reinforce the emerging view that cluster magnetisation and merger-driven turbulence were well established by $z \sim 1$. 

Together, these studies indicate that the parameter space of cluster-scale non-thermal emission is rapidly filling in. The detection fraction of diffuse radio halos at high redshift remains low (10\%), but this primarily reflects current survey depth rather than an intrinsic absence of emission \citep[e.g.,][]{digennaro25}. Deep integrations with MeerKAT and LOFAR now routinely achieve sub-100 $\mu$Jy~beam$^{-1}$ noise, recovering ultra-steep-spectrum ($\alpha \lesssim -1.5$) halos that were previously undetectable. Complementary polarisation and Rotation Measure (RM) studies further suggest that ordered magnetic fields persist out to $z \sim 1$, hinting at early dynamo amplification \citep[e.g.,][]{2023MNRAS.519.5723O, 2025A&A...694A.125L}. Overall, the landscape of high-redshift cluster studies is transitioning from anecdotal detections to statistical sampling. Continued advances in sensitivity and calibration, culminating with SKA-Low and SKA-Mid, will enable the first population-level constraints on magnetic field evolution, turbulent energy injection, and the cosmic history of non-thermal plasma in the ICM \citep[see][for reviews]{Cassano01.2026.SKA, ArpanPal01.2026.SKA}.

\section{Timescales and mechanisms of early magnetic field growth}\label{theroy}

Magnetic fields in galaxy clusters are believed to have originated from very weak ``seed'' fields that were later amplified as clusters evolved \citep[e.g.,][]{ryu12}. These seeds, with strengths of only nG, may have formed in the early Universe through processes like inflation, phase transitions, or Biermann battery effects. Once the ICM became turbulent during structure formation, these weak fields were rapidly amplified by several processes acting in combination \citep[e.g.,][]{sub06}.

\paragraph{Turbulent dynamo:} The turbulent dynamo is widely regarded as the dominant amplification mechanism in galaxy clusters \citep{schekochihin07}. During mergers and accretion, large-scale gas flows generate turbulence that stretches and folds magnetic field lines, increasing their strength. Magneto Hydrodynamic (MHD) simulations show that this process can amplify weak seed fields to micro-Gauss levels within about 1–3 Gyr, consistent with the timescales of cluster evolution \citep{lehle25}. The resulting field is irregular and filamentary, with enhanced strength near shocks, cold fronts, and turbulent regions of the ICM.

\paragraph{Adiabatic compression:} As gas is compressed during gravitational collapse or cluster mergers, the magnetic field is strengthened through flux freezing, following roughly as B $\propto \rho^{2/3}$ \citep{dolag99, donnert18, bruggen20}. This process enhances existing fields locally, particularly in dense cores and shock-compressed regions. However, compression alone cannot explain the observed micro-Gauss fields, suggesting it primarily acts as a secondary amplifier alongside turbulence.

\paragraph{Shock amplification:} Cluster mergers and large-scale accretion generate powerful shock waves that compress and reorient magnetic fields \citep[e.g.,][]{roettiger99, vazza11}. Behind these shocks, turbulence develops that can further amplify the field strength \citep{ryu03}. Such regions are observed as radio relics, tracing the shock fronts at cluster outskirts. Shock compression is particularly important for generating ordered magnetic fields aligned with the relic edges \citep[e.g.,][]{iapichino12}.

\paragraph{AGN and galactic outflows:} Active galactic nuclei (AGN) and star-forming galaxies inject magnetised plasma into the ICM through jets and winds \citep{kronberg08,xu09}. These outflows not only supply magnetic material but also stir the surrounding medium, driving turbulence that aids further amplification. In the early Universe, when AGN activity was more frequent, this process likely played a major role in enriching and magnetising the forming cluster environment \citep{furlanetto01}.

\paragraph{Plasma and kinetic effects:} The ICM is a weakly collisional plasma, where microscopic processes such as mirror and firehose instabilities can modify the effective viscosity and enhance small-scale turbulence \citep{kunz14, bru14}. These effects may increase the efficiency of the dynamo and help sustain magnetic growth in the hot, tenuous gas typical of high-redshift clusters.

\begin{figure*}[t!]
    \centering
    \includegraphics[width= 0.80\textwidth]{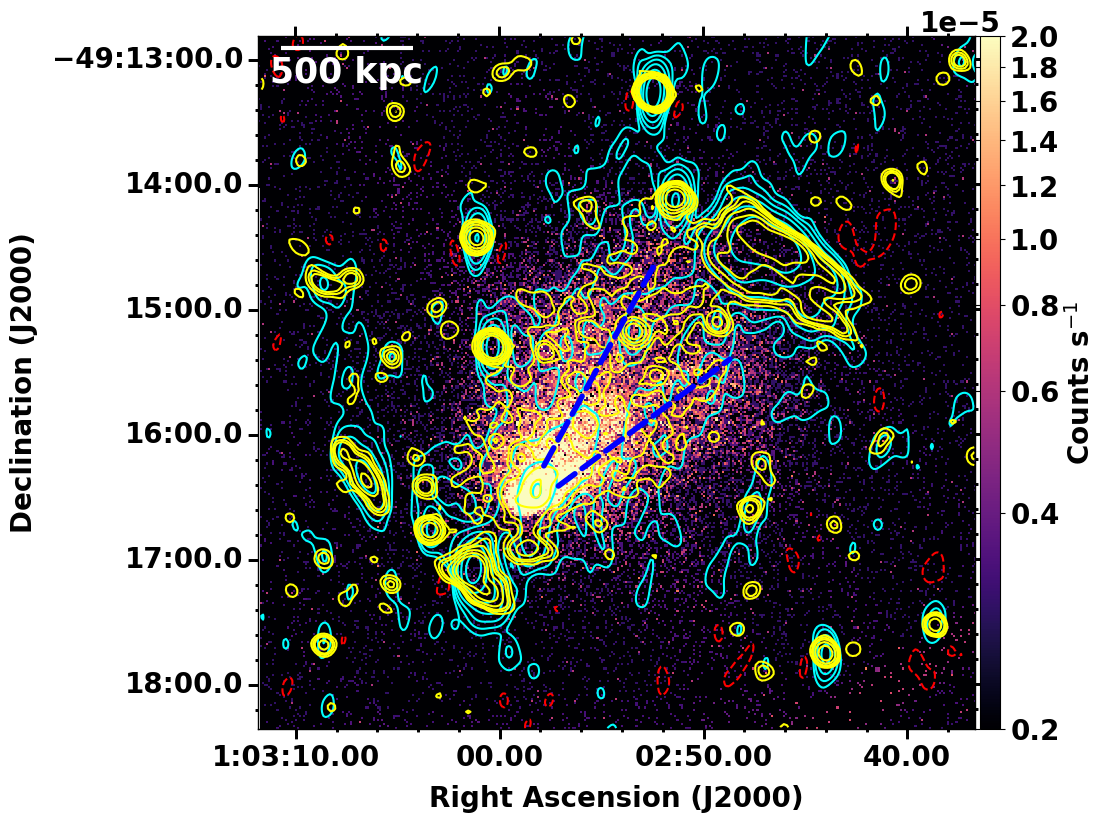}
    \caption{ uGMRT 372 MHz image of El Gordo in cyan (red -ve) contours [-0.1,0.1,0.2,...mJy~beam$^{-1}$] and MeerKAT 1.28 GHz image in yellow contours [15, 30, ... $\mu$Jy~beam$^{-1}$]  on the \emph{Chandra} X-ray image shown in colour. Dashed lines mark the two tails seen in the X-ray. The image is reproduced from \citealt{kale25}.}
    \label{fig:elgordo1}
\end{figure*}

\section{Case Study: The El-Gordo Cluster (ACT-CL J0102$-$4915)} \label{el-gordo}

El Gordo (ACT-CL J0102$-$4915; $z = 0.87$) is one of the most massive and X-ray luminous clusters known at high redshift, undergoing a major merger between two subclusters \citep{2012ApJ...748....7M, Frye_2023}. With a total mass of $2.16\times10^{15}$ M$_\odot$, an intracluster temperature of $>$14 keV, and a strong Sunyaev–Zel’dovich (SZ) signal, it shows a double-peaked mass distribution and elongated X-ray morphology indicative of a dynamically disturbed system \citep[e.g.,][]{2012ApJ...748....7M,2014ApJ...785...20J, 2015ApJ...813..129Z, Kim_2021}. Its discovery demonstrated that such massive systems already existed when the universe was less than half its current age \citep[e.g.,][]{2021MNRAS.500.5249A, Asencio_2023}.

El Gordo hosts a bright northwestern (NW) radio relic discovered with 610 MHz GMRT and 2.1 GHz ATCA observations, along with a faint radio halo and a counter relic in southeast (SE) direction \citep{2014ApJ...786...49L}. The NW relic shows an integrated polarisation fraction of $\sim 33\%$ (at 2.1 GHz), supporting shock-driven Fermi I acceleration. A subsequent 610 MHz GMRT study confirmed the halo and identified a $M\gtrsim3$ shock at the relic location using deep \emph{Chandra} observation \citep{2016MNRAS.463.1534B}. Magnetic field structures were found to align with the merger axis using the synchrotron intensity gradient technique \citep{2024NatCo..15.1006H}. More recent uGMRT observations at 400, 650 and 1400 MHz, combined with MeerKAT 1.28 GHz and \emph{Chandra} data, provided spatially resolved spectral index maps and turbulence estimates \citep{kale25}. The diffuse radio components are designated as RH, NW-relic, SE-relic, E-relic and ERelic-Ext (Figure~\ref{fig:elgordo1}).

El Gordo’s extreme merger energy makes it a powerful testbed to study magnetic-field amplification and cosmic-ray acceleration at high redshift. At $z = 0.87$, inverse Compton losses are significant, offering a stringent test of turbulent dynamo efficiency. Its high merger energy provides strong turbulence and shock-driven flows that can be directly compared with predictions from cosmological MHD simulations \citep[e.g.,][]{2015ApJ...813..129Z, ng2015}. The system has also been extensively observed across the electromagnetic spectrum, from \emph{Chandra} and XMM-Newton in X-rays, to HST in the optical, ACT and SPT in SZ, and several radio arrays, making it uniquely suited for cross-validation of thermal and non-thermal correlation studies.

The SKAO will enable a transformative leap in the study of El Gordo and similar high-redshift systems. With sub-arcsecond resolution and few $\mu$Jy sensitivity, SKA-Low and SKA-Mid will resolve the morphology and spectral curvature of the diffuse emission, distinguish embedded radio galaxies from genuine ICM structures, and measure Faraday rotation with unprecedented precision \citep[e.g.,][]{2020Galax...8...53H}. These data will allow mapping of magnetic-field coherence scales, polarisation fractions, and spectral ageing, providing direct constraints on field topology and acceleration mechanisms. El Gordo thus serves as a benchmark system for the high-redshift cluster population that the SKA will unveil—linking present-day cluster magnetism to its origins during the epoch of large-scale structure formation.

\section{Prospects with the SKA AA* and AA4}\label{ska}

The upcoming SKA AA* (Array Assembly) and AA4 configurations will enable high dynamic range imaging of diffuse radio structures beyond Mpc-scales. In particular, the extended mid-baseline coverage in the AA4 configuration reaching up to $\sim$159.6 km will significantly enhance angular resolution, achieving $\sim$3$''$ at 200 MHz and $\sim$6$''$ around 75 MHz, offering images comparable in resolution to that of the GMRT at 610 MHz and LOFAR at 110 MHz, but with improved \textit{uv}-coverage. The combination of SKA-Low and SKA-Mid will revolutionise our ability to study the non-thermal ICM in distant clusters such as El Gordo, delivering unprecedented sensitivity, resolution, and polarimetric precision.

At low frequencies (50$-$350 MHz), SKA-Low will be optimally suited for detecting the steep-spectrum synchrotron emission from ageing cosmic-ray electrons in cluster halos and relics. In its full AA4 configuration, SKA-Low will reach continuum sensitivities of 5$-$10 $\mu$Jy~beam$^{-1}$ for deep (10$-$20 hr) integrations with 3$-$8$^{\prime\prime}$ resolution, corresponding to $\sim$21$-$50 kpc at z $\sim$ 0.87. These observations will recover the diffuse, low-surface-brightness emission that current facilities (LOFAR, uGMRT, MeerKAT) only marginally detect (Table \ref{tab:ska_sensitivity}). The large collecting area and dense core of the array will allow sensitivity to angular scales exceeding 10$'$ at high redshift, enabling full mapping of the extended radio halo and peripheral relics of El Gordo. SKA-Low will thus directly trace the large-scale distribution of magnetised plasma and re-acceleration regions, even up to the cluster virial radius \citep[e.g.][]{Cuciti01.2026.SKA}, constraining the total magnetic energy density and the efficiency of merger-driven turbulence at early cosmic epochs.

At higher frequencies (0.35$-$15 GHz), SKA-Mid will complement this picture by probing the morphological, spectral, and magnetic structure of the non-thermal emission at sub-arcsecond resolution. With rms sensitivities of $\sim$1$-$2 $\mu$Jy~beam$^{-1}$ in 10 hr and $\sim$0.1 $\mu$Jy~beam$^{-1}$ in 50 hr integrations, SKA-Mid will detect emission down to a few $\mu$Jy~arcsec$^{-2}$ when smoothed to 5–10$^{\prime\prime}$ resolution (Table \ref{tab:ska_sensitivity}). This corresponds to sub $\mu$G level magnetic-field sensitivity in the ICM. Such depths will allow detection of diffuse structures with surface brightnesses as low as 0.05$-$0.5 $\mu$Jy~arcsec$^{-2}$ with signal to noise $>$ 10, and enable polarisation detections for fractional polarisation levels of 5\% on 5–10$^{\prime\prime}$ scales. The wide instantaneous bandwidth of SKA-Mid will provide in-band spectral-index mapping, revealing curvature and breaks due to electron ageing, while high-fidelity polarisation and rotation-measure (RM) synthesis will measure field coherence lengths and strengths with precisions of $\lesssim 5$ rad~m$^{-2}$.

\begin{figure}
    \centering
    \includegraphics[width=\textwidth]{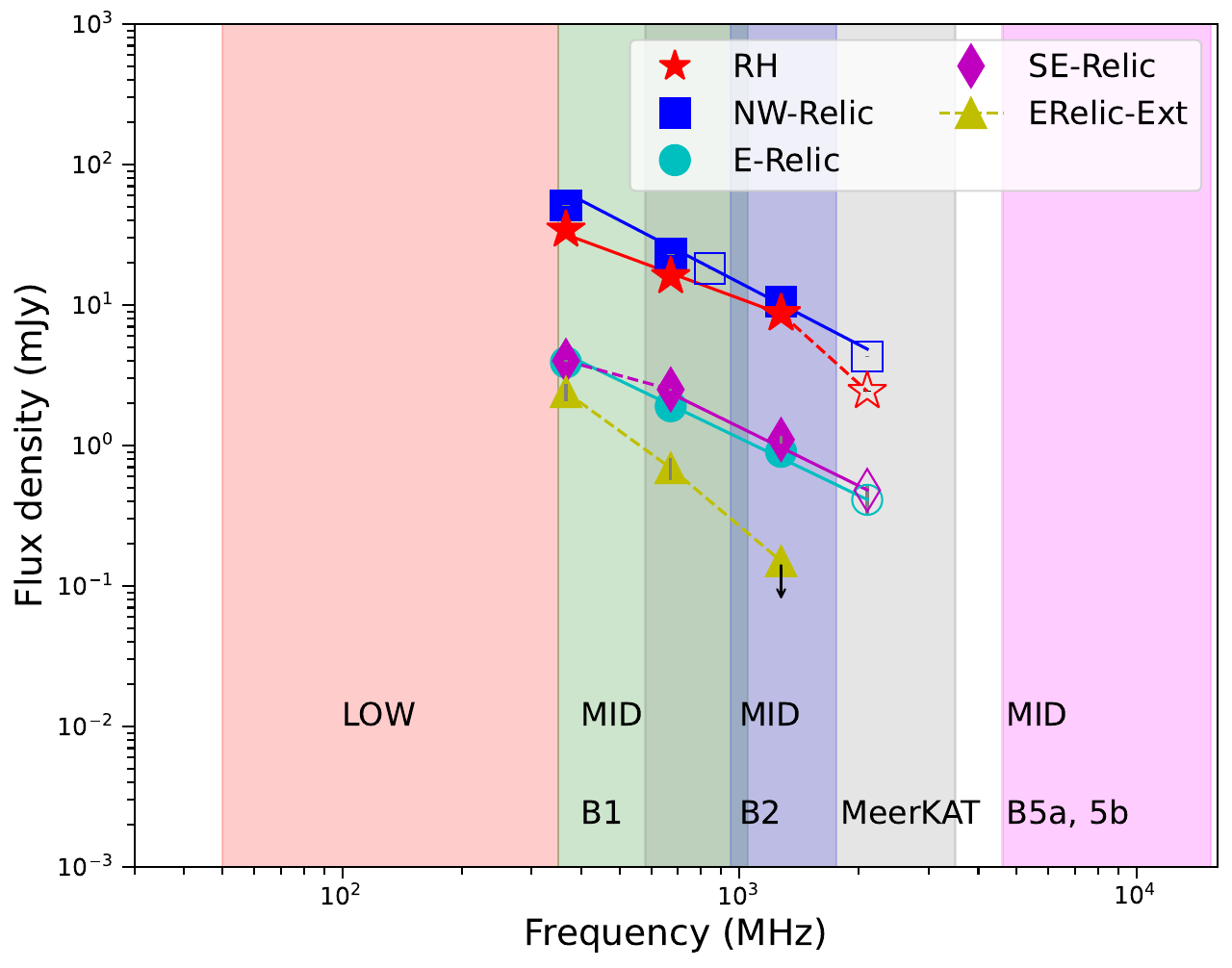}
    \caption{The schematic integrated spectrum for the different diffuse structures in the El-Gordo is shown with SKA-Low and SKA-Mid bands specified. The flux density values for each of the diffuse sources are taken from the \citealt{kale25}.}
    \label{fig:schematic_intspec}
\end{figure}

In practice, SKA-Low will define the global morphology and detect the faintest halo envelope, while SKA-Mid will dissect the internal structure, polarisation, and spectral gradients of the emission. Together they will offer a complete, multi-frequency view (Figure.~\ref{fig:schematic_intspec}) of El Gordo—from the large-scale diffuse halo tracing merger-driven turbulence to the sharp relic edges delineating shock acceleration. The combined dataset will, for the first time, permit direct reconstruction of the magnetic-field topology, cosmic-ray ageing, and energy coupling between thermal and non-thermal components in a high-redshift cluster. Through such deep observations, SKA will not only transform our understanding of El Gordo but also establish the observational framework for studying the cosmic evolution of magnetised plasma in the first few billion years of the universe.

\begin{table}[ht!]
\centering
\caption{Comparison of sensitivities and observational parameters between uGMRT and SKA AA* configurations.}
\label{tab:ska_sensitivity}
\renewcommand{\arraystretch}{1.2}
\begin{tabular}{lccccc}
\hline
Configuration 
& Freq 
& Resolution 
& RMS noise 
& SB sensitivity 
& Confusion noise \\

& GHz & $''$ & $\mu$Jy~beam$^{-1}$ & K & $\mu$Jy~beam$^{-1}$ \\

\hline
uGMRT                & 0.25--0.5  & 6--8      & 25.00 & 1.06 & 2.00 \\
SKA AA* (Low)       & 0.05--0.35 & 8         & 8.66  & 6.07 & 8.35 \\
SKA AA* (Mid)       & 0.35--1.05 & 2--3      & 1.72  & 0.65 & 0.33 \\
SKA AA* (Band 2)    & 0.95--1.76 & 1--2      & 0.83  & 0.16 & 0.047 \\
SKA AA* (Band 5a)   & 4.6--8.5   & 0.1--0.2  & 0.34  & 0.18 & 0.00 \\
SKA AA* (Band 5b)   & 8.3--15.4  & 0.07--0.08 & 0.41 & 0.21 & 0.00 \\
\hline
\end{tabular}
\end{table}

\section{Sensitivity and detectability using SKA}\label{population}

To date, only a handful of massive systems, such as El\,Gordo at $z \sim 0.87$, have double relic and halo with highly extended emission, limiting statistical studies of cluster diffuse emission in this untapped regime. Upcoming SKA surveys, however, will dramatically change this landscape, enabling a population-level investigation of high-redshift clusters. \cite{cassano15} provided an approach for estimating the detectability of radio halos as a function of redshift, and we have given some estimation on the 3$\sigma$ limit at SKA-Low frequencies. The minimum detectable radio power $P_{\nu, \rm min}(z)$ at a frequency $\nu$ can be approximated as:
\begin{equation}
P_{\nu, \rm min}(z) \simeq 4 \pi D_L^2(z) \, S_{\rm rms} \, f(\theta_H, z),
\label{eq:Pmin}
\end{equation}
where $D_L(z)$ is the luminosity distance at redshift $z$, $S_{\rm rms}$ is the survey sensitivity (in W~m$^{-2}$~Hz$^{-1}$), $\theta_H$ is the angular size of the radio halo, and $f(\theta_H, z)$ accounts for the dilution of surface brightness over the beam. For typical SKA-Low and SKA-Mid surveys, $S_{\rm rms} \sim 5~\mu$Jy~beam$^{-1}$ and $1~\mu$Jy~beam$^{-1}$, respectively, with angular sizes of $\sim 1$~Mpc assumed for halos at high redshift. This equation allows one to compute the minimum radio power detectable as a function of redshift, which can then be directly compared with observed cluster powers. Applying this framework to El\,Gordo, the observed 150~MHz flux densities of $S_{150} = 42$~mJy, correspond to a rest-frame 150~MHz radio power of 6.05 $\times 10^{25}~\rm W~Hz^{-1}$, assuming a typical spectral index of $\alpha = -1.3$. This positions El\,Gordo well above the nominal SKA1-LOW detection limit at $z=0.87$, making it a suitable testbed for probing high-redshift ICM physics.

\begin{figure}
    \centering
    \includegraphics[width=0.49\textwidth]{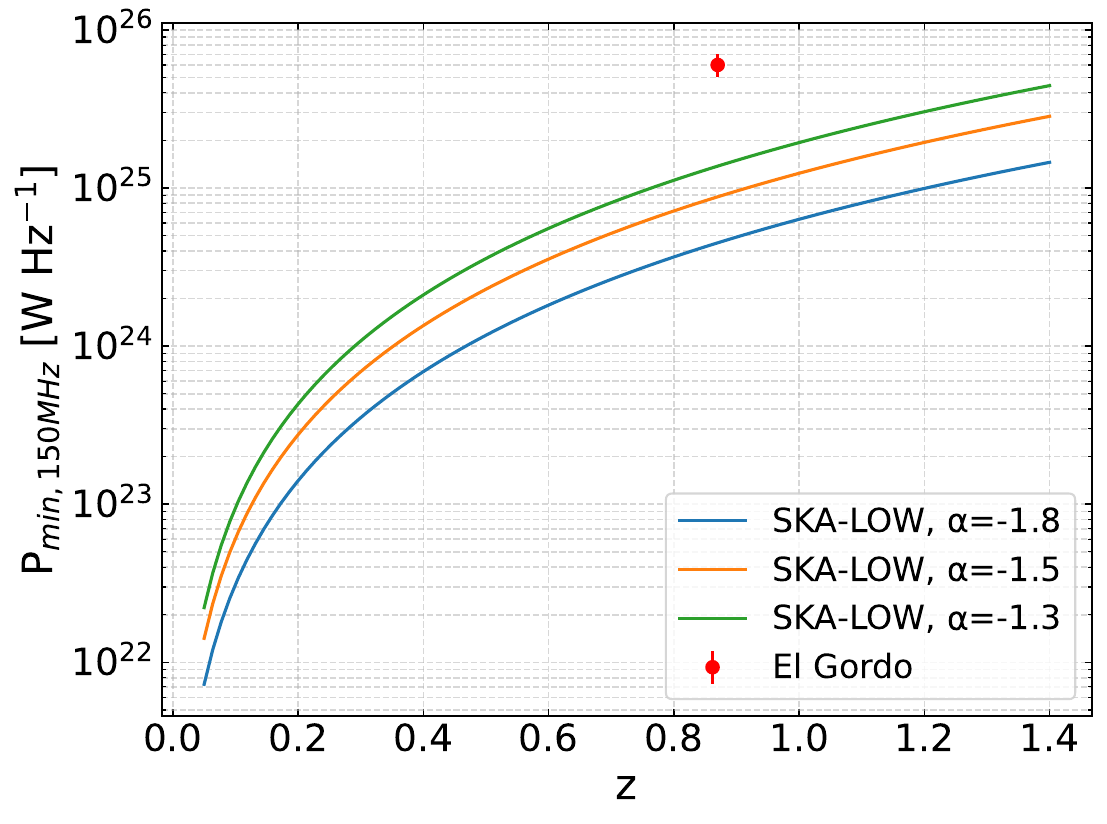}
    \includegraphics[width=0.49\textwidth]{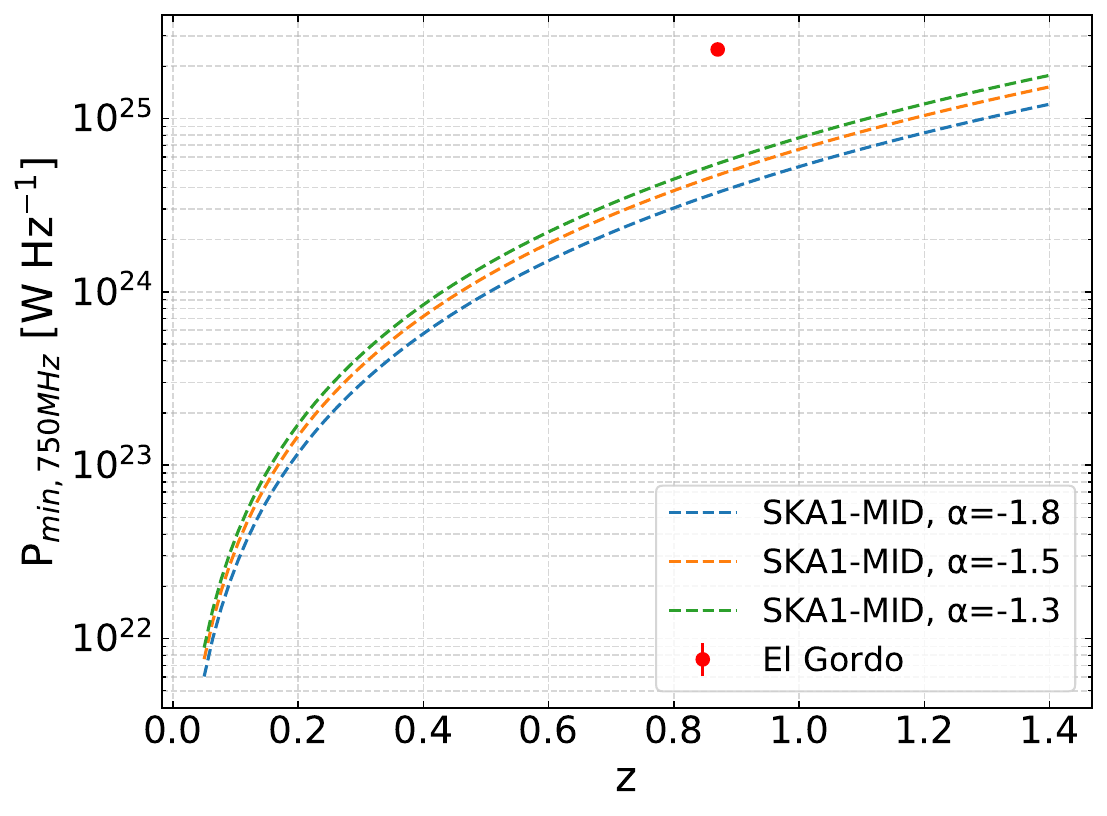}
    \caption{\textit{Left:} Minimum power of RHs detectable at 150 MHz as a function of redshift is shown. The solid line indicates the detectability for different spectral indices. The minimum radio power has been computed according to Eqs.~\ref{eq:Pmin}. \textit{Right:} The same is shown here at the SKA-MID frequency, with the same spectral index values (each dotted curve).}
    \label{fig:schematic_min_dectection}
\end{figure}

The sensitivity curves (shown in Figure~\ref{fig:schematic_min_dectection}) derived from Equation~\ref{eq:Pmin} demonstrate that SKA-LOW ($\sim 100-350$~MHz) is particularly sensitive to steep-spectrum halos with $\alpha \gtrsim 1.3$, enabling the detection of systems with $P_{1.4\,\rm GHz} \gtrsim 3\times10^{24}~\rm W~Hz^{-1}$ at $z \sim 1$. SKA-MID ($\sim 0.5-1.5$~GHz) complements this capability by providing higher-resolution imaging and improved spectral constraints, although it is somewhat less sensitive to faint, steep-spectrum halos at high redshift. These estimations indicate that a statistically meaningful sample of extended emission in several hundred high-redshift clusters should be detected. Such population studies will allow direct investigation of the evolution of the $P_{1.4\,\rm GHz}$--$M_{500}$ scaling relation and the growth of magnetic fields and turbulence in the ICM over cosmic time. A key challenge for SKA-Low is that observations approach the confusion limit rapidly, making the separation of diffuse emission from compact or extended radio galaxies difficult. In the next section, we will discuss this issue and also quantify the population-level expectations in the SKA era.

\section{Expectations on high-redshift cluster population}\label{sec:pop}

\begin{figure}
    \centering
    \includegraphics[width=0.80\textwidth]{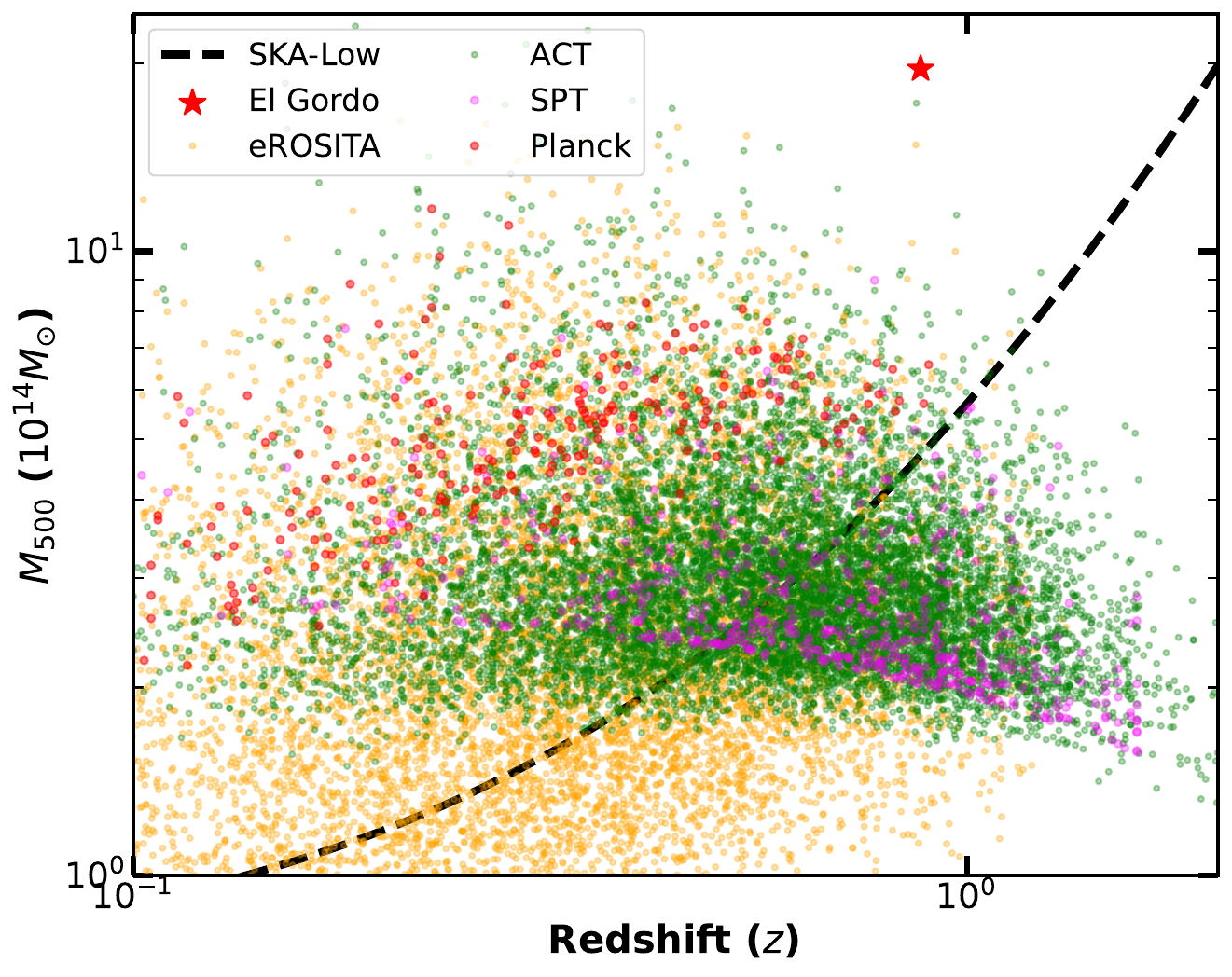}
    \caption{Cluster mass–redshift distribution illustrating the discovery space for diffuse radio emission studies. The plotted plane shows $M_{500}$ (in units of $10^{14}M_\odot$) as a function of redshift, highlighting the parameter space covered by major high-redshift multi-wavelength surveys in the X-ray and SZ \citep{bulbul24, bocquet19}. The location of El Gordo is marked for reference. The black line indicates the sensitivity regime accessible to SKA-Low AA4 (assuming the redshift dimming effect), demonstrating its capability to probe lower-mass clusters at progressively higher redshifts compared to current facilities.}
    \label{fig:m-z}
\end{figure}

El Gordo lies at the extreme high-mass end of the known high-redshift cluster distribution, well above the $M_{500}\sim(3-8)\times10^{14}$M$_\odot$ range that dominates current $z>0.6$ samples (Figure~\ref{fig:m-z}). Therefore, understanding the radio luminosity, spectral index, and surface-brightness structure in such a system will help to anchor the high-mass end of the scaling relations that SKA will test across hundreds of clusters. This calibration is essential for interpreting whether fainter halos in lower-mass $z\sim0.8-1$ systems represent scaled-down versions of extreme mergers like El Gordo or a fundamentally different regime of non-thermal activity.

In LOFAR 144 MHz observation for 8 hr, reaching $\sim200~\mu{\rm Jy~beam^{-1}}$ at $\sim20''$ resolution, halos at $z\sim0.6$ can only be detected above $P_{\rm 150~MHz}\gtrsim1.5\times10^{24}~{\rm W~Hz^{-1}}$ \citep{digennaro25}, with the detection threshold increasing by nearly an order of magnitude by $z\sim1.5$. Ultra deep ($\sim 50 - 100$ hr) integrations on individual targets can lower the noise; however, such efforts are not practical for the large high-redshift samples from the South Pole Telescope \citep[SPT;][]{bocquet19} and the Atacama Cosmology Telescope \citep[ACT;][]{calabrese25}. At high redshift, isolating diffuse halo emission from cluster radio galaxies also becomes increasingly difficult. Using different {\it uv}-ranges (i.e., restricting the baseline lengths in the Fourier plane) can suppress compact sources, but when a halo spans only a few beams, this approach quickly loses effectiveness, and extended, or remnant radio galaxies can blend seamlessly with the diffuse emission \citep[e.g.][]{kale25}.

SKA-Low will markedly improve the situation, reaching $\sim20~\mu{\rm Jy~beam^{-1}}$ at $\sim10''$ resolution in just one hour at 150 MHz and opening access to much fainter halos. Even so, SKA-Low observations will approach the confusion limit at low resolution. For this reason, coordination with SKA-Mid is essential: its sharper resolution and reduced confusion will allow accurate modelling and subtraction of embedded radio galaxies, enabling reliable measurements of the underlying halo population. The SKA-Low telescope, particularly in its AA4 configuration (with 74 km of baseline), will provide an unprecedented view of diffuse non-thermal emission from galaxy clusters across the southern sky. This part includes the richest compilation of massive clusters currently known, thanks to recent and ongoing X-ray and SZ surveys such as \textit{eROSITA}, \textit{SPT}, and \textit{ACT} (Figure~\ref{fig:m-z}). These surveys span a wide region of the M$_{500}$ vs $z$ plane, extending to high redshift ($z \sim 1.5$) and probing the most massive systems in the Universe.

As shown in Fig.~\ref{fig:m-z}, SKA-Low surveys in AA4 configuration will have access to a large fraction of these clusters, specifically: more than 500 at ($z \gtrsim 0.6$), above a mass range of 5 $\times$ 10$^{14}$M$_{\odot}$. A population of southern clusters occupies a parameter space in mass and redshift that remains largely unexplored at low radio frequencies. Ongoing efforts such as the MeerKAT Massive Distant Clusters Survey (MMDCS) are already extending this approach to systematically probe the non-thermal properties of high-redshift clusters \citep{phuravhathu25}. In particular, SKA-Low surveys will: (i) build a statistically significant populations of high-redshift system $z>0.6$; (ii) test long-standing particle acceleration models from merger shocks and turbulence in a cosmologically evolving environment; (iii) probe how rapidly magnetic fields are amplified during cluster assembly stage, and verify some of the models discussed in section~\ref{theroy}, and (iv) reveal the connection between cluster mass assembly and the emergence of relativistic plasma and magnetic fields in the intracluster medium.

\section{Summary}\label{summ}

High-redshift clusters provide a powerful window into the early growth of magnetic fields in the intracluster medium. Over the past decade, observations with LOFAR, uGMRT, and MeerKAT have detected diffuse radio halos and relics out to $z \sim 1$, showing that $\mu$G-level magnetic fields and efficient cosmic-ray acceleration were already in place when the universe was only a few billion years old. These results challenge classical dynamo timescales and instead suggest rapid magnetic-field amplification driven by mergers, turbulence, and shocks, operating efficiently enough to offset strong inverse Compton losses. El Gordo (ACT-CL J0102$-$4915, $z = 0.87$), one of the most massive and dynamically disturbed clusters at high redshift, hosts a bright relic, a radio halo, and additional diffuse structures linked to a violent merger. Its extreme mass and well-studied multi-wavelength properties make it a benchmark system for investigating magnetic amplification under conditions of strong turbulence and enhanced inverse Compton losses.

The SKA will substantially advance this field by combining sensitivity, resolution, and survey scale. In its AA4 configuration, SKA-Low will reach $\sim20~\mu{\rm Jy~beam^{-1}}$ in one hour at 150 MHz with $\sim10''$ resolution, and 5–10 $\mu$Jy~beam$^{-1}$ in deep integrations, enabling the detection of ultra-steep-spectrum halos well beyond the current LOFAR limits. Its dense core will recover diffuse emission on scales exceeding 10$'$, allowing full mapping of Mpc-scale halos and peripheral relics even at $z \gtrsim 1$. SKA-Mid will complement this capability with sub-arcsecond resolution and $\sim1$–2 $\mu$Jy~beam$^{-1}$ sensitivity in 10-hour integrations, providing in-band spectral-index mapping, accurate rotation measures ($\lesssim5$ rad m$^{-2}$), and polarisation constraints at a few percent level. Together, the AA* and AA4 configurations will deliver the combination of surface-brightness sensitivity and angular resolution required to understand the complex interplay of shock and turbulence in the ICM.

Beyond such testbed systems as El Gordo, SKA surveys will access more than $\sim$500 clusters at $z \gtrsim 0.6$ with $M_{500} \gtrsim 5\times10^{14},M_\odot$, transforming high-redshift galaxy cluster studies. This will allow direct measurements of the evolution of the $P_{1.4,\rm GHz}$–$M_{500}$ relation, the occurrence fraction of diffuse emission, and the efficiency of particle re-acceleration under enhanced inverse Compton losses. By comparing radio power, morphology, and polarisation properties with predictions from cosmological MHD simulations, SKA will place quantitative limits on magnetic seed-field scenarios and the timescale of turbulent amplification. In doing so, it will close the long-standing observational gap at $z>0.6$ and provide robust constraints on the growth of magnetised plasma during the first few billion years of the galaxy cluster assembly.

\bibliographystyle{abbrvnat-maxbibnames4}
\bibliography{chapter} 

\end{document}